\documentclass[journal]{IEEEtran}
\usepackage{cite}
\usepackage[pdftex]{graphicx}
\graphicspath{{Images/}}
\DeclareGraphicsExtensions{.pdf,.jpeg,.png,.eps}
\usepackage{epstopdf}
\usepackage{amsmath}
\usepackage{algorithmic}
\usepackage{array}
\usepackage{graphics}
\usepackage{adjustbox}
\usepackage{fixltx2e}
\usepackage{stfloats}
\usepackage{url}

\usepackage{booktabs}
\usepackage{longtable}
\usepackage{multicol}
\usepackage{caption}
\usepackage{multirow}
\usepackage{rotating}
\usepackage{float}
\usepackage{flushend}
\usepackage[referable]{threeparttablex}
\usepackage{xtab}
\usepackage{ragged2e}
\usepackage{fancyhdr}
\makeatletter
\def\ps@IEEEtitlepagestyle{%
  \def\@oddhead{\mycopyrightnotice}%
  \def\@evenhead{}%
}
\def\mycopyrightnotice{%
  {ASIAN JOURNAL OF ENGINEERING, SCIENCES \& TECHNOLOGY, VOL.5, ISSUE 1 \hfill MARCH 2015}
  \gdef\mycopyrightnotice{}
}
\fancypagestyle{plain}{
  \fancyhf{}
  \fancyhead[L]{ASIAN JOURNAL OF ENGINEERING, SCIENCES \& TECHNOLOGY, VOL.5, ISSUE 1 \hfill MARCH 2015}     
    \fancyfoot[C]{\thepage}

}
\begin{document}
\title{Design and Implementation of a DTMF Based Pick and Place Robotic Arm}
\author{Muhammad Hassan and Mohtashim Baqar\thanks{Muhammad Hassan was with the Department of Telecommunication Engineering, IQRA University (Main Campus), Karachi (e-mail: hasan.ovais@gmail.com).}\thanks{Mohtashim Baqar is with the Faculty of Engineering, Sciences and Technology, IQRA University (Main Campus), Karachi, Pakistan (e-mail: mohtashim@iqra.edu.pk).}}
\maketitle
\pagestyle{plain}
\begin{abstract}
In recent times, developments in field of communication and robotics has progressed with leaps and bounds. In addition, the blend of both disciplines has contributed heavily in making human life easier and better. So in this work while making use of both the aforementioned technologies, a procedure for design and implementation of a mobile operated mechanical arm is proposed, that is, the proposed arm will be operated via a cellular device that connects with the receiver mounted on the robotic arm. Moreover, over the duration of a call, if any key is pressed from the cellular device than an indicator indistinct to the key pressed is noticed at the receiver side. This tone represents superimposition of two distinct frequencies and referred to as DTMF (dual tone multi-frequency). Further, the mechanical arm is handled via the DTMF tone. Also, the acquired tone at the receiver is taken into a micro-controller (ATMEGA16) using the DTMF decipher module i.e. MT8870. Further, the decipher module unwinds the DTMF signal into its corresponding two bit representation and then the matched number is transmitted to the micro-controller. The micro-controller is programmed to take an action based on the decoded value. Further, the micro-controller forwards control signals to the motor driver unit to move the arm in forward/backward or multi-directional course. Lastly, the mechanical arm is capable of picking and placing objects while being controlled wirelessly over GSM (Global System for Mobile Communications).
\end{abstract}
\IEEEpeerreviewmaketitle
\section{Introduction}
Robotics has been an area which has seen a lot of growth and development in last 15 years \cite{balasubramanian2009object}. Today, a big part of resources i.e. man power or money etc., allocated for research have been utilized for development in the said discipline as it holds a lot of applications in both commercial and military settings \cite{wong1994phone,coskun1998remote,moutinho2003progresses}. Recently, a lot of the developments have been driven by the motivation of it to be used for the betterment of the society and to reduce human effort or workload. Robot is an electro-mechanical machine and an automated electro-mechanical vehicle can help humans to operate at places where physical presence is difficult or impossible because of the circumstances  i.e. space or place set on fire etc. In this a work a procedure for the implementation of DTMF operated robotic arm is proposed for picking and placing of objects. Further, the earlier customary remote controlled arms have constraints, such as, weak human interface, circuit weaknesses, conducive environment for remote operation and restricted control. Nonetheless, these constraints may be minimized by utilizing the wireless innovations as a part of the electro-mechanical arm \cite{sai2009design,sharma2006dtmf,mustafa2012vehicle}. Further, doing so would add further more to the applications, strength and usability of the arm. In this work, DTMF has been used for operating the mechanical arm wirelessly over GSM \cite{aranguren2002remote}. Moreover, dual tone multiple frequency (DTMF)  signalling has been utilized for telecom motioning over simple telephony lines. The said is achieved while using the voice recurrence bands among telephonic sets and different specialized gadgets. Further, the variants of DTMF utilized for telephone tone calling are renowned and a term Touch-Tone is used to represent the process. Also, it has been standardized by ITU-T as recommendation Q.23. It is also referred to as MF4 in the Great Britain. Further, other multi-recurrence frameworks are also available utilizing the in-band signalling for different control purposes. Also, in early 80's, DTMF tones were likewise used by high quality TV telecasters to demonstrate the starting and halting times of station breaks. \par 
Dual tone multiple frequency \cite{gulledge1997automated} has been the foundation of tone of voice articulation management. Existing day telephonic connections make use of DTMF in order to dial volumes, configure telephone trades (switchboards) coming from distant control destinations and so forth. \par
\begin{figure}[H]
\includegraphics[width=\columnwidth]{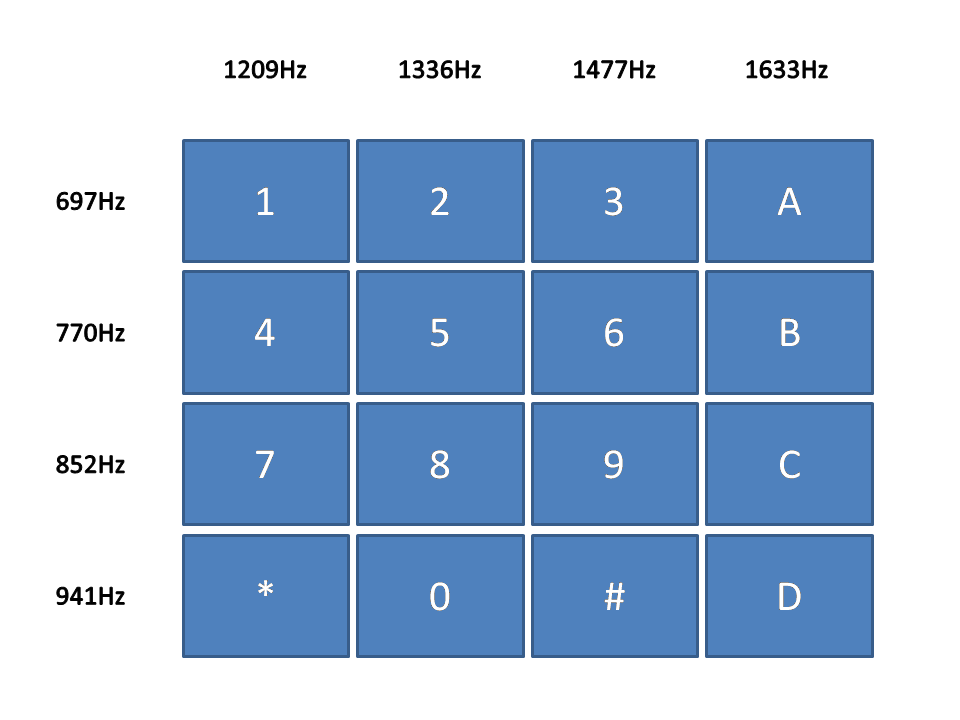}
\caption{Dual Tone Multiple Frequency Map (DTMF)}
\label{figure:DTMF}
\end{figure}

 
Figure \ref{figure:DTMF} illustrates dual tone multiple frequency (DTMF) out map in form of a $4\times 4$-array keypad. It gives an illustration of the dual-tone frequency combinations generated on respective key presses on the keypad.
Further, A, B, C and D keys are not usually utilized and conventional keypads don't have these keys. However, these keys were earlier used a lot in land line telephone and fax systems to configure with the local exchanges or to perform specialized tasks. When a key is pressed on the keypad, a tone, combination of two superimposed frequencies, is generated and send to the local exchange for decoding. These combinations of frequencies were utilized to issue instructions and operate the electro-mechanical arm wirelessly. 
\section{Connection Layout}
\begin{figure}[H]
\centering
\includegraphics[width=\columnwidth]{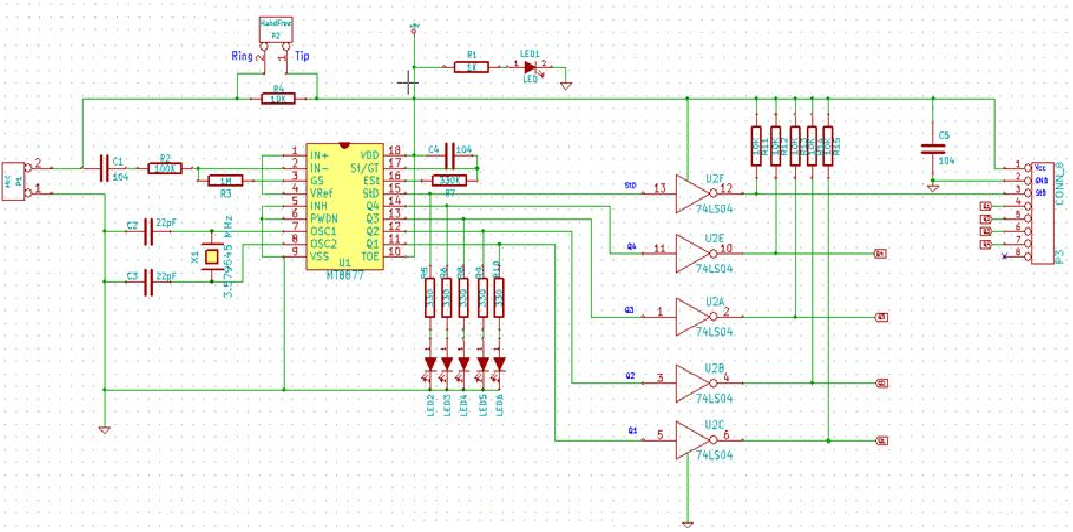}
\caption{Receiver - DTMF Interfacing Circuit}
\label{figure:DTMF_IN}
\end{figure}

Fundamental building blocks of the electro-mechanical arm are the DTMF decoder, micro-controller, motor driver circuity and power supply unit. A MT8870 DTMF decoder has been utilized in this work. Further, different versions of MT8870 are available utilizing the advanced digital counting techniques to detect and decode all the 16 DTMF tone sets into their respective 4-bit codes. Moreover, there wasn't any need of a preprocessing or pre-filtering circuit for the removal artefacts from the generated dual-tone signal. This is due to the built-in noise rejection filter in the signal generating source, mitigating the adverse effects of any unwanted frequency component. Figure \ref{figure:DTMF_IN} illustrates the circuit diagram for the DTMF decoder. 
The fundamental blocks of this automated mechanical arm are as follows,
\begin{enumerate}
\item DTMF Decoder
\item ATMEGA32 Micro-controller 
\item Motor Driver 
\item Regulated Power Supply 
\end{enumerate}
Portion below gives a brief overview of the fundamental modules of the complete model.
\subsection{DTMF Decoder}
\begin{figure}[H]
\centering
\includegraphics[width=2.5in]{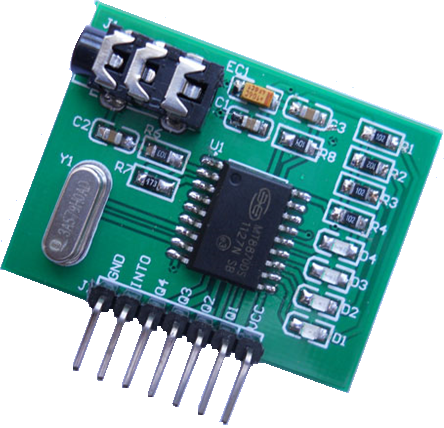}
\caption{DTMF Decoder - MT8990}
\label{figure:MT8990}
\end{figure}

\begin{figure}[H]
\centering
\includegraphics[width=2.5in]{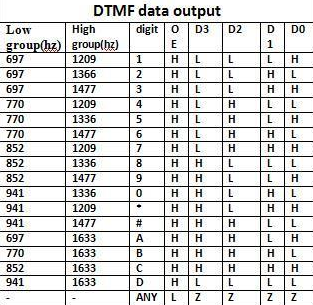}
\caption{DTMF - Output Map}
\label{figure:DO}
\end{figure}

In this work, a MT8870 arrangement DTMF decipher is utilized. The MT8870d/ MT8870d-1 is an accomplished DTMF collector having the advanced decipher capacity. The channel segment uses twisted capacitor systems for higher and lower gathering channels and the decoder utilizes advanced checking procedures to identify and translates every one of the 16 DTMF tone sets into a 4-bit code. Different varieties of MT8870 arrangements utilize the computerized matching systems to discover and discriminate all the 16 DTMF tone sets into a 4-bit code. Figure \ref{figure:DO} and \ref{figure:MT8990} illustrates the DTMF decoder and output map of the DTMF decoder, respectively.
\subsection{ATMEGA-32 Micro-controller}
In this work, an ATMEL micro-controller has been used. The ATMEL AVR module has 32 registers. Further, the said 32 registers are connected with the arithmetic logic unit (ALU), letting two different registers to be worked on a single clock cycle. The design is more effective while accomplishing cycles up to ten times quicker than true CISC micro-controllers.
\subsection{L293 Motor Driver}
For motor driving, L293 motor driver has been used \cite{singh1991some}. L293 is an incorporated circuit motor driver that could be utilized for multi and bidirectional navigation of two motor. Moreover, it's load driving limit is till 600 m-amp, however, it can only handle a less amount of current compared to it's maximum value due to excessive rise in temperature during regular operations. \par
Further, L293 is available normally in a 16-pin package, a dual-in-line incorporated circuit bundle. Moreover, there is another variant of it with the name, L293D. The "D" adaptation is picked in the greater part of the cases as it has implicit fly back diodes that minimize the inductive voltage spikes. \par
\begin{figure}[H]
\centering
\includegraphics[width=2.5in]{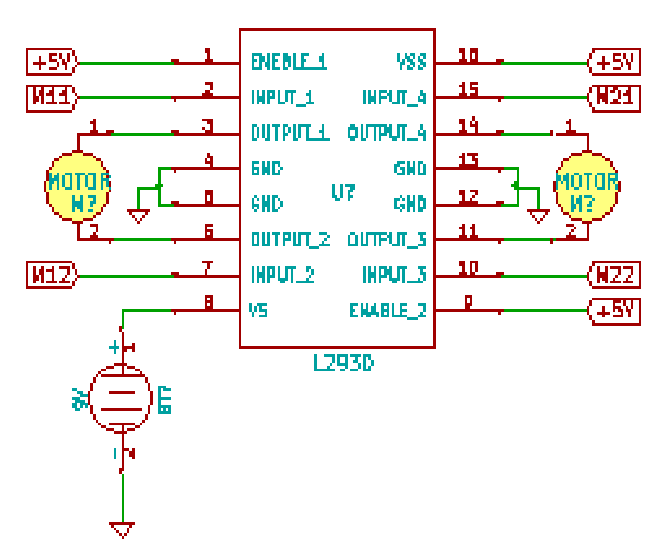}
\caption{Motor Driving Circuitry}
\label{figure:MD}
\end{figure}

Further, two DC motors and two servo motors have been used in this implementation to drive the base on which the arm has been mounted and to move the arm, respectively. Figure \ref{figure:MD} illustrate circuit diagram of the motor driving unit.
\subsection{Power Supply}  
A switch-controlled power supply has been used in this work \cite{smallhorn2006seat}. It is one in which the produced voltage is constantly balanced as per the necessities. Further, the operating voltage range of the power supply used in this work is $\pm 36$ volts with an available current of $12 amp$..
\section{Reverse Kinematics}
Kinematics is a field describing motion of an object or structure. It has two main parts namely, forward kinematics and reverse kinematics. \par 
Forward kinematics uses kinematic equations of a robot to find the position of end-effectors from joints inside the structure.\par
\begin{figure}[H]
\centering
\includegraphics[width=\columnwidth]{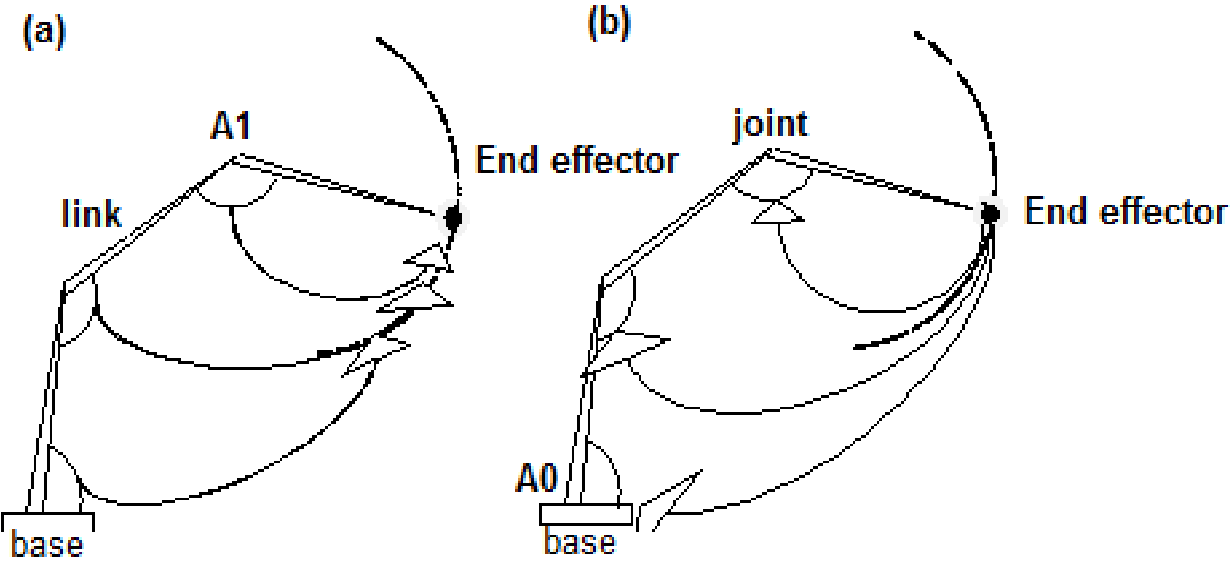}
\caption{Structure Manipulation \textbf{(a)} Forward \textbf{(b)} Inverse}
\label{figure:EF}
\end{figure}

 
But if one is willing to familiarize that how the upper joint framework would keep on motion when it is obligatory that the end effectors should move in the direction of its mean position. In that case direct estimations are worked out in the mechanical framework so that the connecting angle can be calculated as the alterations are done in the end effectors. Figure \ref{figure:EF} illustrates structure manipulation via forward and inverse kinematics. 
Therefore, it can be said that the notion of reverse kinematics employs joint parameter estimation with the help of kinematic equations involving robotics while the desired position of the end effectors is known \cite{sciavicco1996modeling}.
\section{MATHEMATICAL ANALYSIS}
There could be many desirable outcomes if only the mathematical values of the end effectors are given as the input. So for avoiding repetition, one more gripper angle is taken with respect to the horizontal as input. \par
\begin{figure}[H]
\centering
\includegraphics[width=2.3in]{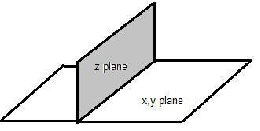}
\caption{XYZ Plane}
\label{figure:PL}
\end{figure}

 
These are supposed as the input to a robotic arm:
\begin{enumerate}
\item The x, y, z coordinate of the end effectors. 
\item The following values should be known: base length (1), shoulder length (2), arm length (3), gripper length (4) 
\item Gripper Angle (g) 
\item Radial Length (r) 
\end{enumerate}
For ease, the estimations of 3-D inverse kinematics are changed into 2-D form. For this purpose, two planes are being examined to be rectilinear with each other: z-plane and x, y-plane in horizontal as shown in figure \ref{figure:PL}.
Moreover, mathematical outcomes in the x, y plane use coordinates x and y. \par 
Further, the position of the z-plan is determined with the help of the rotating motor as it goes across the coordinates (0, 0, 0) and the radial distance (r) is given at (x, y, 0) coordinates.
\section{CALCULATED TORQUE}
Torque is the measure of how many forces are acting on an object which in result make that object to rotate. It is denoted by T. Torque (T) is termed as a rotating "force" and is estimated using the following equation.
\begin{equation}
T=F*L
\end{equation}
where, T represents torque, F represents calculated force and L denotes the length from the pivot point. Also, force is the acceleration of an object due to gravity (g = 9.81 $m/s^2$) multiplied by its mass.
\begin{equation}
F=M*g 
\end{equation}
where, 'M' denotes mass and 'g' denotes gravity.\par
\begin{figure}[H]
\centering
\includegraphics[width=2.3in]{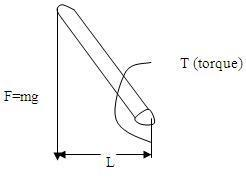}
\caption{Torque Balance}
\label{figure:TA}
\end{figure}

 
In addition, force (F) is also considered to be equal to an object's weight(W). Mathematically,
\begin{equation}
W = M*g
\end{equation}
Therefore, the torque needed to grab a mass at a given displacement from a pivotal point is given by,
\begin{equation}
T = (M*g)*L
\end{equation}
Where, the length 'L', is the perpendicular length from a pivot point to the force. Similarly, length can also be found by doing a torque balance about a point. Mathematically,
\begin{equation}
\sum T = 0 = F*L – T
\end{equation}
Hence, replacing the force (F) with mass and gravity (m*g), we can find out the same equation above. This is the more accurate way to find out the torque by using the torque balance. Figure \ref{figure:TA} illustrates torque balancing. Mathematically,
\begin{equation}
M*g*L = TA
\end{equation}
\section{Sample Working}
\begin{figure}[H]
\centering
\includegraphics[width=2.3in]{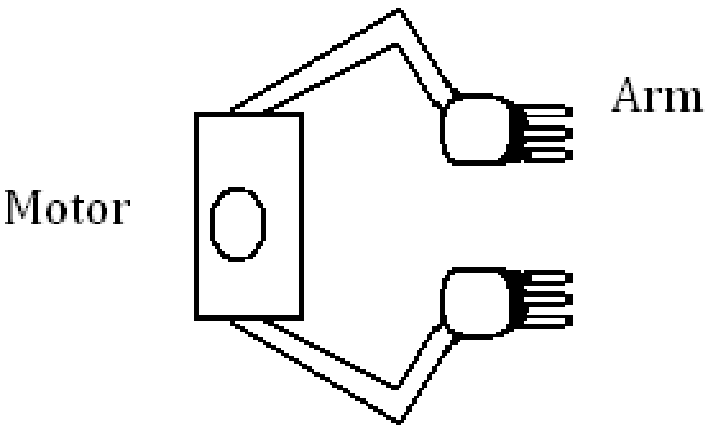}
\caption{Torque Balance}
\label{figure:Gripper}
\end{figure}

The implemented electro-mechanical arm can lift up to 10 kg of weight. Figure \ref{figure:Gripper} illustrate the gripper of the robotic arm. Further, servo motors were used for strengthening and loosening of grip of the gripper as well as for the movement of the arm whereas DC gear motors were used to move the arm in forward and backward direction. With key presses using the key pad of any cellular device one can operate the robotic arm subject to that the cellular device is connected to the one mounted on the robotic arm.
\section{APPLICATIONS AND FUTURE SCOPE}
As already mentioned that robotics have greater application in military and rather in all walks of life. Examples being the unmanned aerial vehicles, capturing an militants, spying and rescuing people. Quite a few nations are contemplating to make utilization of these vehicles in case of any catastrophes \& natural calamity. Also, these automated vehicles will also be helpful to reach and access places where human reach and presence is not possible under the circumstances.
\section{Acknowledgement}
The authors would like to acknowledge the Faculty of Engineering, Sciences and Technology (FEST), IQRA University for their persistent support and encouragement throughout the course of the project.
\section{Conclusion}
In this implementation, an electro-mechanical arm controlled via cellular phone has been developed that makes a call to the cell phone attached to the arm. During the duration of the call, if any key is pushed, a tone mapped to the key pushed is acquired at the receiver side. This remotely operated electro-mechanical arm eliminates the constraint of a cabled medium via use of wireless technologies, that is, GSM and DTMF in this case. \par
Further, work can still be done to enhance the strength and capacity of this framework. Cell phone that makes a call to cellular phone attached to the base of the mechanical arm provides remote access and henceforth, this does not need the development of receiving and transmitting units and also it is free from the issues related to RF communication and it's reachability. Moreover, this model could be extremely useful in events of data collection from areas of human outreach, which has been one of the prime causes of rise in research activities in this domain.
\bibliographystyle{IEEEtran}
\bibliography{ref}
\begin{IEEEbiographynophoto}{Muhammad Hassan}
received the B.E degree in Telecommunication from IQRA University (Main Campus), Karachi 2013, respectively.
\end{IEEEbiographynophoto}
\vspace{-0.4in}
\begin{IEEEbiographynophoto}{Mohtashim Baqar}
received the B.E and M.S degrees in Telecommunication and Electrical Engineering from IQRA University (Main Campus), Karachi and National University of Sciences and Technology, Islamabad, Pakistan, in 2011 and 2016, respectively. Since 2012, he has been a Lecturer in Faculty of Engineering, Sciences and Technology, IQRA University (Main Campus), Karachi. His research interests are multi-variate signal processing and pattern recognition.
\end{IEEEbiographynophoto}
\end{document}